\documentclass[nofootinbib,preprint,superscriptaddress,amsmath,amssymb,aps,pra]{revtex4}

\usepackage{amsmath,amsfonts,amssymb}
\usepackage{graphicx}
\usepackage{xcolor}
\usepackage{float}

\usepackage{calc}

\usepackage{accents}

\begin{document}

\title{Deep learning for retrieval of the internuclear distance in a molecule from interference patterns in photoelectron momentum distributions}

\author{N. I. Shvetsov-Shilovski}
\email{n79@narod.ru}
\affiliation{Institut f\"{u}r Theoretische Physik, Leibniz Universit\"{a}t Hannover, 30167 Hannover, Germany}

\author{M. Lein}
\affiliation{Institut f\"{u}r Theoretische Physik, Leibniz Universit\"{a}t Hannover, 30167 Hannover, Germany}

\date{\today}

\begin{abstract}
We use a convolutional neural network to retrieve the internuclear distance in the two-dimensional H$_2^{+}$ molecule ionized by a strong few-cycle laser pulse based on the photoelectron momentum distribution. We show that a neural network trained on a relatively small dataset consisting of a few thousand of images can predict the internuclear distance with an absolute error less than $0.1$~a.u. We study the effect of focal averaging, and we find that the convolutional neural network trained using the focal averaged electron momentum distributions also shows a good performance in reconstructing the internuclear distance.\\

% \noindent Discipline: Atomic, Molecular and Optical, Facet: Research Areas, Concept: Multiphoton or tunneling ionization and excitation.\\
% Facet: Physical Systems, Concept: Atoms.\\
% Facet: Techniques, Concepts: Deep Learning, Time-Dependent Schr\"odinger equation
 
\end{abstract}

%\pacs{32.80Fb, 32.80Rm, 32.80 Wr}

%\keywords{strong-field physics}

\maketitle

\newpage

\section{Introduction} Development of techniques aimed at visualization of electronic and molecular dynamics in real time will open new horizons in many branches of modern science and technology, including chemistry, biology, and material science. Many different techniques for time-resolved molecular imaging have been proposed thus far (see Ref.~\cite{Agostini2016} for a review). These methods have emerged as a result of remarkable progress in laser technologies over the last decades. The emergence and availability of table-top intense femtosecond laser systems has led to several new time-resolved imaging techniques using the highly nonlinear phenomena originating from interaction of strong laser pulses with atoms and molecules (see, e.g., Refs.~\cite{Graefe2016, Lin2018} for review of these phenomena and the whole field of strong-field physics). Examples are laser-assisted electron diffraction \cite{Kanya2010,Morimoto2014}, laser induced Coulomb explosion imaging \cite{Frasinski1987,Cornaggia1991,Posthumus1995,Cornaggia1995}, high-order harmonic orbital tomography \cite{Itatani2004,Haessler2010}, laser-induced electron diffraction (LIED) \cite{Lein2002,Meckel2008,Blaga2012,Pullen2015}, and strong-field photoelectron holography (SFPH) \cite{Huismans2011}.  The two latter methods analyze momentum distributions of electrons from strong-field ionization. The recent experimental achievements in LIED and SFPH (see, e.g., Refs.~\cite{Haertelt2016,Walt2017}) suggest that future experiments will aim at extracting the information about nuclear motion in a molecule from electron momentum distributions. 

The understanding of the outcomes of these forthcoming experiments requires thorough theoretical studies of the effects of nuclear motion on the photoelectron momentum distributions. Such theoretical studies are already on the way. For instance, it was shown in Ref.~\cite{Haertelt2016} that the different nuclear wave packet dynamics in hydrogen and deuterium molecules leads to a difference in bond length, which, in turn, transforms into a shift of the holographic fringe at certain electron momenta. Before analyzing the imprints of the nuclear motion in momentum distributions, it is useful to study the distributions for fixed nuclei with varying of internuclear distance. In the present paper we address this problem using methods of machine learning, which is ``a subfield of Computer Science wherein machines learn to perform tasks for which they were not explicitly programmed" \cite{TraskBook}.  

Machine learning, and more specifically deep learning (see, e.g., Refs.~\cite{RashkaBook,KimBook} for textbook treatments), has been successfully applied to a number of problems in strong-field physics and adjacent fields of research. This includes the prediction of the flux of high-order harmonics for different experimental parameters using a neural network \cite{Gherman2018}, the prediction of the ground state energy of an electron in various 2D confining potentials by training a convolutional neural network (CNN) \cite{Mills2017}, and efficient numerical implementation of the trajectory-based Coulomb-corrected strong-field approximation (TCSFA) (see Refs.~\cite{Yan2010,Yan2012} for the foundations of the TCSFA method) with a deep neural network \cite{Yang2020}. In all these examples the application of machine learning allowed to avoid heavy computational costs that would be inevitable when solving these problems using traditional ways.

Very recently the CNN's were used to predict high-order harmonic generation (HHG) spectra for model di- and triatomic molecules for randomly chosen parameters. The latter include laser intensity, internuclear distance, and orientation of the molecule \cite{Lytova2021}. Furthermore, it was shown in Ref.~\cite{Lytova2021} that the CNN can be used for solving the inverse problems: determination of molecular and laser parameters, as well as classification of molecules based on their HHG spectrum (or time-dependent dipole acceleration) alone.  These problems are hard to solve by manually inspecting a variety of complex spectra. On the other hand, classification is one of the typical tasks of machine learning. A similar situation is found for the problem of the present work. Indeed, identification of the changes in photoelectron momentum distributions with varying internuclear distance in a molecule is a complicated task. This problem can be solved by applying machine learning. In this paper we train a CNN to predict the internuclear distance in the two-dimensional (2D) model H$_{2}^{+}$ molecule from a given photoelectron momentum distribution (PMD). The momentum distributions that are needed to train the CNN are calculated from the direct numerical solution of the time-dependent Schr\"{o}dinger equation (TDSE). We show that a good accuracy of the predictions can be achieved even for relatively small sets of training data. We then study the effect of the focal averaging on the retrieval of the internuclear distance with the CNN. 

The paper is organized as follows. In Sec. II we discuss our approach to numerical solution of the 2D TDSE and sketch the architecture of the used CNN. In Sec.~III we apply the neural network to predict the internuclear distance in the H$_2^{+}$ molecule. Then we discuss the effect of focal averaging on the retrieval of the distance between the nuclei. The conclusions and outlook are given in Sec.~IV. 

\section{Model}

\subsection{Solution of time-dependent Schr\"odinger equation}

For the calculations we use a few-cycle linearly polarized laser pulse that is defined in terms of the vector-potential and present between $t=0$ and $t_f=\left(2\pi/\omega\right)n_p$:
\begin{equation}
\label{vecpot}
\vec{A}\left(t\right)=\left(-1\right)^{n_p}\frac{F_0}{\omega}\sin^2\left(\frac{\omega t}{2n_p}\right)\sin\left(\omega t+\varphi\right)\vec{e}_{x}.
\end{equation}
Here $\vec{e}_x$ is a unit vector in the polarization direction ($x$-axis), $n_p$ is the number of optical cycles within the pulse, and $\varphi$ is the carrier-envelope phase (CEP). The electric field is to be obtained from Eq.~(\ref{vecpot}) as $\vec{F}\left(t\right)=-d\vec{A}/dt$. We do our simulations for $\varphi=0$. The (positive-valued) pre-factor $\left(-1\right)^{n_p}$ in Eq.~(\ref{vecpot}) ensures that for $\varphi=0$ the electric field component $F_x\left(t\right)$ has its maximum at the center of the pulse, i.e., for $\omega t=\pi n_p$.

In the velocity gauge, the 2D TDSE for an electron interacting with the laser pulse is given by 
\begin{equation}
\label{2d_tdse}
i\frac{\partial}{\partial t}\Psi\left(x,y,t\right)=\left\{-\frac{1}{2}\left(\frac{\partial}{\partial x^2}+\frac{\partial}{\partial y^2}\right)-iA_{x}\left(t\right)\frac{\partial}{\partial x}+V\left(x,y\right)\right\}\Psi\left(x,y,t\right),
\end{equation}
where $\Psi\left(x,y,t\right)$ is the coordinate-space wave function and $V\left(x,y\right)$ is the soft-core binding potential of the model H$_{2}^{+}$ molecular ion in the frozen nuclei approximation:
\begin{equation}
\label{potential}
V\left(x,y\right)=-\frac{1}{\sqrt{\left(x-R/2\right)^2+y^2+a}}-\frac{1}{\sqrt{\left(x+R/2\right)^2+y^2+a}}.
\end{equation}
Here, $R$ is the internuclear distance and $a=0.64$ is the soft-core parameter. The corresponding time-independent Schr\"odinger equation for the bound states reads as
\begin{equation}
\label{tise}
\left\{-\frac{1}{2}\left(\frac{\partial}{\partial x^2}+\frac{\partial}{\partial y^2}\right)+V\left(x,y\right)\right\}\Psi\left(x,y\right)=E\Psi\left(x,y\right).
\end{equation}
We use the Feit-Fleck-Steiger split-operator method \cite{Feit1982} to solve the TDSE Eq.~(\ref{2d_tdse}). The ground-state wave function was obtained by imaginary time propagation. We find that, for instance, for $R=3.0$ a.u. the energy of the ground state in the potential given by Eq.~(\ref{potential}) is $E_0=-0.915$~a.u. Our computational box was centered at $\left(x=0,y=0\right)$ and extends over $x\in\left[-400,400\right]$~a.u. and $y\in\left[-200,200\right]$~a.u. We use equal grid spacings for $x$ and $y$ coordinates, $\Delta x=\Delta y=0.1954$~a.u., corresponding to 4096 and 2048 points in $x$- and $y$-direction, respectively. 

The wave function was propagated from the beginning of the laser pulse $t=0$ to $t=4t_f$ with the time step $\Delta t=0.0184$~a.u. We apply absorbing boundaries to prevent unphysical reflections of the wave packet from the boundary of the computational grid, i.e., at every time step the wave function is multiplied by the mask:
\begin{equation}
M(x,y) = 
 \begin{cases}
   1 &\text{for $r\leq r_b$}\\
   \exp\left[-\beta\left(r-r_b\right)^2\right] & \text{for $r > r_b$}
 \end{cases}
\end{equation}
Here, $r=\sqrt{x^2+y^2}$, $r_b=150$~a.u. and $\beta=10^{-4}$. We note that at the intensity of $4.0\times10^{14}$ W/cm$^2$ and for the wavelength of 800~nm, the characteristic amplitude of the laser-induced electron quiver motion is $F_0/\omega^2=32.8$~a.u. Therefore, the position of the absorbing boundary $r_b$ exceeds this value by a factor of $4.5$. The photoelectron momentum distributions are calculated by using the mask method \cite{Lein2002,Tong2006}. 

\subsection{Architecture of convolutional neural network}

The choice of architecture of a neural network should account for the structure of the data used for learning and the desired output. In our case the data used for learning are the pairs consisting of the PMD (image) and the corresponding internuclear distance $R$ (label). Bearing in mind that the H$_2^{+}$ molecule is ionized by a laser field linearly polarized along the internuclear axis, we assume that every PMD has the aspect ratio $2:1$. To be more specific, we work with grayscale images of the size of $256\times128$ pixels. The details of the necessary image preprocessing are given in Sec.~III. We train the neural network to solve the regression problem, i.e., to predict the internuclear distance $R$ from a given PMD. 

\begin{figure}[h!]
\begin{center}
\includegraphics[trim=0 50 0 0, width=.80\textwidth]{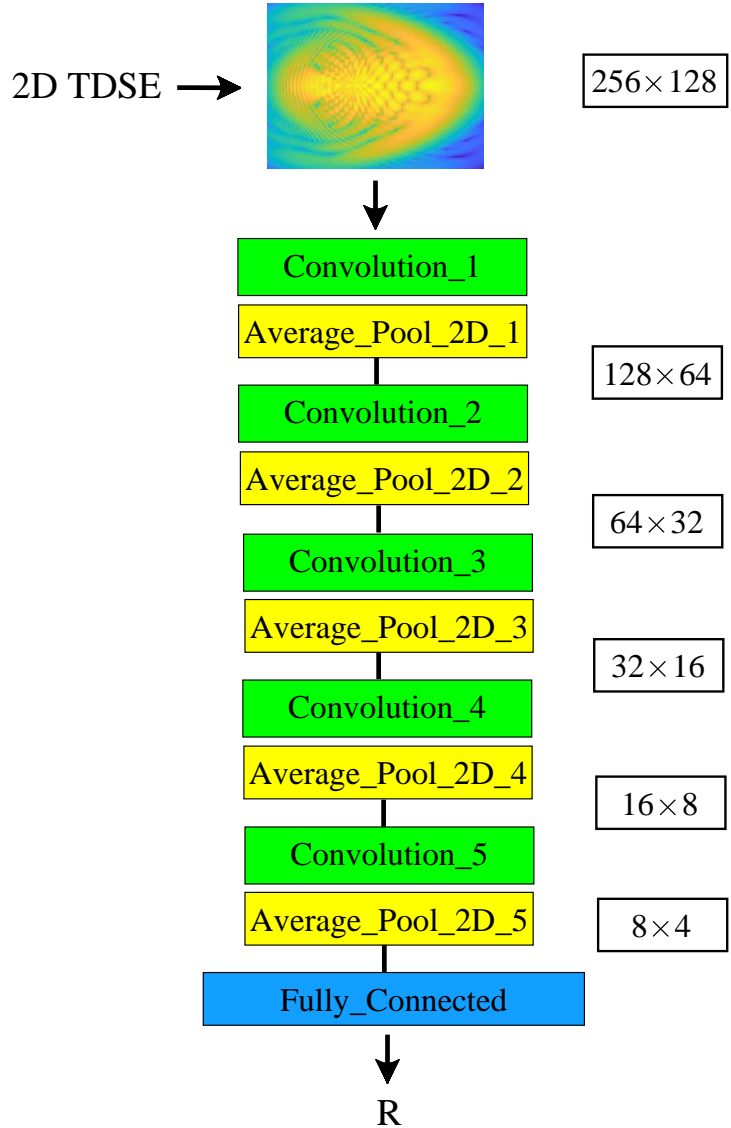} 
\end{center}
\caption{The architecture of the neural network used for retrieval of the internuclear distance $R$. As an input, the CNN receives the momentum distribution of size of $256\times128$ pixels calculated from the solution of the 2D TDSE. The neural network consists of 5 pairs of layers, each composed of convolutional layer with a ReLU activation function and an average pooling layer. The sizes of the image after each pair of layers are indicated on the right. The fully connected layer at the end acts as a decision layer that predicts the internuclear distance.}
\label{fig1}
\end{figure}

The architecture of the deep neural network that we use for the problem at hand is shown in Fig.~\ref{fig1}. The neural network consists of $5$ nonreducing convolutional layers, each followed by a reducing average pooling layer. Each of the nonreducing convolutional layers operates with 32 filters with sizes of $3\times3$ pixels. These convolution layers produce new images ``feature maps" (see, e.g., Ref.~\cite{KimBook}). The number of these new images equals the number of filters. The values of the filter matrices are to be determined through the training process, i.e., they play the same role as the trainable weights of an ordinary artificial neural network. After performing the convolution operation, all the convolutional layers apply the rectified linear unit (ReLU) activation function, which is defined as $\textrm{ReLU}(x)=\max(0,x)$. The average pooling layers divide the images they get into pooling regions with sizes of $2\times2$ pixels and calculate averaged values in every region. Therefore, each average pooling layer reduces the size of the image by a factor of $2$. The last average pooling layer is connected to the dropout layer (not shown in Fig.~\ref{fig1}) that randomly sets its input elements, i.e., output of the preceding layer, to zero with a certain probability. This probability is chosen to be equal to $0.2$. The dropout layer allows to avoid overfitting. The output of the dropout layer is fed to a fully connected layer that produces only one single value: the internuclear distance $R$. This is the output value of the whole neural network.

\section{Numerical experiments and results}

\begin{figure}[H]
\begin{center}
\includegraphics[width=.90\textwidth, trim=0 30 0 0]{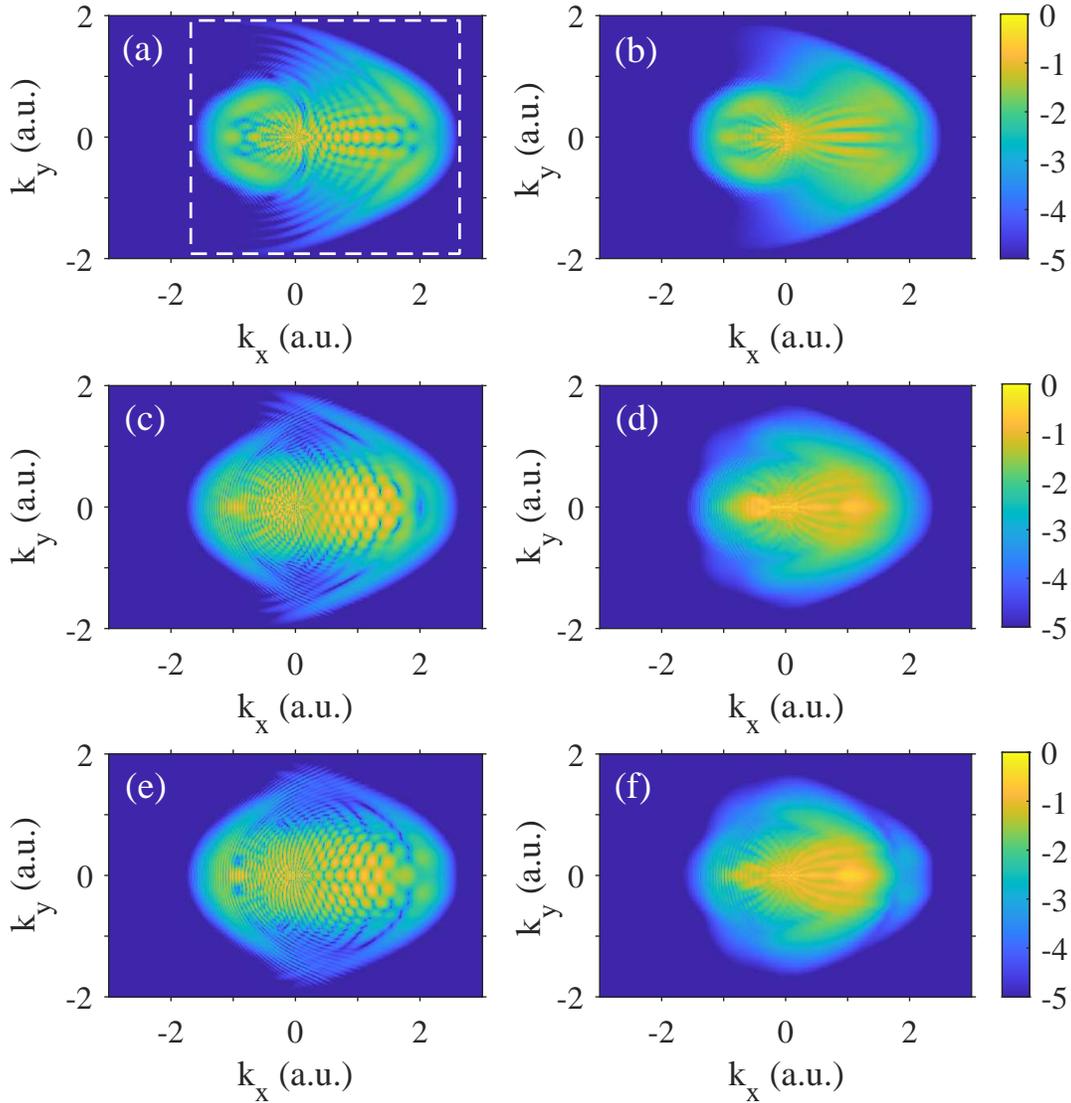} 
\end{center}
\caption{Electron momentum distributions for ionization of the H$_2^{+}$ molecule by a laser pulse with a duration of $n_p=2$ cycles and wavelength of $\lambda=800$~nm obtained from numerical solution of the TDSE. The panels (a, b), (c, d), and (e, f) correspond to the internuclear distances 2.0~a.u., 5.0~a.u., and 6.0~a.u., respectively. The left column [panels (a), (c), and (e)] show the distributions calculated at fixed intensity of $\text{4.0}\times\text{10}^{\text{14}}$~W/cm$^2$. The right column [panels (b), (d), and (f)] displays the distributions averaged over the focal volume for the same peak intensity of $\text{4.0}\times\text{10}^{\text{14}}$~W/cm$^2$. The laser field is linearly polarized along the $x$-axis. The distributions are normalized to the maximum value. Shown is the decimal logarithm of the distribution, see text.}
\label{fig2}
\end{figure}

The photoelectron momentum distributions calculated from the solution of the TDSE (\ref{2d_tdse}) for three different internuclear distances are shown in Figs.~\ref{fig2}~(a), (c), and (e). It is seen that the shape of the distribution changes considerably with increasing $R$. However, the quantification of the corresponding changes in the PMD’s is a non-trivial task. This makes application of neural networks particularly appropriate.

In order to train the neural network, we first need to produce a set of training data. To this end, we solve the TDSE, Eq.~(\ref{2d_tdse}), for $N_{\text{train}}$ random internuclear distances $R_k\in\left[1.0, 8.0\right]$~a.u. and peak laser intensities $I_{k}\in\left[1.0, 4.0\right]\times 10^{14}$ W/cm$^2$, where $k=1,...,N_{\text{train}}$, and we calculate the corresponding electron momentum distributions. Since the solution of the 2D TDSE takes a few hours on 4 to 8 modern CPU's working in parallel, the formation of a large training set is computationally expensive. Here we use $N_{\text{train}}=3000$. We note that our data set is relatively small (compared with $N_{\text{train}}=200000$ and $N_{\text{train}}=30000$ used in Refs.~\cite{Mills2017} and \cite{Lytova2021}, respectively). Nevertheless, even such a modest data set allows us to obtain satisfying results. 

The PMD calculated from the solution of the TDSE is a matrix of size $4096\times2048$. The usage of matrices of such sizes as an input for a convolutional neural network will lead to a slow training process. Indeed, convolution of a large input is a computationally expensive operation, and convolutional layers perform a lot of such operations in each forward and backward pass of training images through the neural network. For this reason, we first modify the matrix of the PMD as follows. 

% \begin{figure}[h]
% \begin{center}
% \includegraphics[width=.85\textwidth]{Fig33.eps} 
% \end{center}
% \caption{Preprocessing of the electron momentum distribution. (a) Initial PMD calculated for the intensity of $2.0\times10^{14}$ and $R=2$. The rest of parameters are the same as in Fig.~2. The rectangular region, where the probability exceeds $10^{-5}$ of the maximum of the distribution, is shown by dashed white lines. The solid white lines pass through the centers of some interference maxima. (b) The rectangular region of panel (a), which is resized to $256\times128$ pixels by using bicubic interpolation.}
% \label{fig3}
% \end{figure}

We find the absolute maximum $\text{PMD}_{\text{max}}$ of the distribution and calculate the decimal logarithm of the normalized PMD: $W=\log_{10}\left(\text{PMD}/\text{PMD}_{\text{max}}\right)$. We set $W=-5$ for all values that are smaller than -5. Therefore, we account only for the values exceeding $10^{-5}\ \text{PMD}_{\text{max}}$. We note that in doing so we consider not only the low-energy part of the distribution created by the electrons that do not experience hard recollisions with their parent ions, but also the beginning of the high-energy part of the PMD. This high-energy part is formed due to electrons that are driven back by the laser field to their parent ions and rescatter from them. Classically, the boundary between low- and high-energy parts of the PMD corresponds to the momentum $k=2\sqrt{U_p}$, where $U_p$ is the ponderomotive potential. For the parameters of Fig.~\ref{fig2}, this estimate yields $k\approx1.87$~a.u. Then we find a rectangular area such that the values of $W$ at the boundary of the rectangle are just above -5. This rectange is shown by the dashed lines in Fig.~\ref{fig2}~(a). The image withing the rectangle is resized to $256\times128$ by using bicubic interpolation, which is an extension of a cubic Hermite spline for interpolation on a 2D grid. Finally, all the elements of the matrix are rescaled so that the minimum value corresponds to zero and the maximum one is mapped to 255. The resulting matrix of the size of $256\times128$ is used as an input for the CNN.  

We split our data set into training and test sets in the ratio 0.75:0.25. Only data from the training set were used for training of the neural network. The goal of training is the minimization of the loss function, i.e., the measure of deviation between predictions of the neural network and expected outcomes for the training set. We use the mean squared error as the loss function and apply stochastic gradient descent optimization with momentum. We find that $30$ epochs of training are enough for convergence of the loss function. In order to ensure that each training image creates an unbiased change in the model, we shuffle the training data before each training epoch. The MATLAB package \cite{Matlab} is used for the calculations.

\begin{figure}[h]
\begin{center}
\includegraphics[width=.95\textwidth]{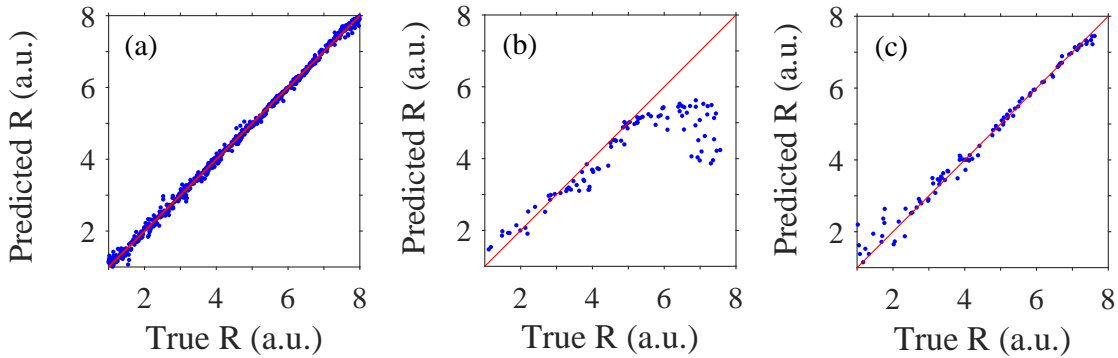} 
\end{center}
\caption{Plot of predicted vs. true internuclear distances illustrating the performance of neural networks. (a) Neural network trained on a set of distributions for fixed laser intensities. (b) The same neural network as in (a), but receiving focal averaged momentum distributions as test images. (c) The neural network  trained on a set of focal averaged distributions and tested on another independent set of focal averaged distributions.}
\label{fig4}
\end{figure}

The results of the application of the trained neural network to the test data are presented in Figs.~\ref{fig4}~(a), where we show the predicted and true values of $R$. It is clearly seen that the neural network can successfully predict the internuclear distance. We note that similar figures were used in Ref.~\cite{Mills2017} to illustrate the performance of the neural network predicting the ground state energy for different confining potentials. We characterize the quality of the neural network by the mean absolute error (MAE) between the predicted and true values of $R$ over the test data set - a measure, which is different from the loss function (mean squared error) used in the training process. The neural network predicts the internuclear distance with the MAE of $0.07$~a.u. We have found that another neural network that uses solely the low-energy part of the PMD's ($0<k_x<\sqrt{2U_p}$, $\left|k_y\right|<\sqrt{U_p}$) shows slightly worse results: the corresponding MAE of $R$ is equal to $0.12$~a.u. This implies that the recognition of the internuclear distance with the neural network relies mostly on the interference patterns in the low-energy part of the momentum distributions, i.e., on the holographic patterns \cite{Huismans2011}, but the accuracy can be enhanced by including high-energy electrons. 

It is clear that the shape of the PMD's depends not only on $R$, but also on the laser parameters, especially on the intensity that fluctuates in an experiment. This raises the question: How vulnerable is the performance of the trained neural network to the effect of focal averaging? To answer this question, we calculate a number of electron momentum distributions averaged over the focal volume and use them to test our neural network trained on the distributions obtained for \textit{fixed} laser intensities. For a peak intensity $I_0$, the focal-volume averaged distribution $dP/d\vec{k}$ can be calculated as \cite{Morishita2007}
\begin{equation}
\label{aver}
\frac{dP}{d^{3}k}=\int_{0}^{I_0}\frac{dP\left(I\right)}{d^{3}k}\left(-\frac{\partial V}{\partial I}\right)dI,
\end{equation}
where $dP\left(I\right)/d^{3}k$ is the momentum distribution for a fixed intensity $I$, and $\left(\partial V/ \partial I\right)dI$ is the focal volume element that corresponds to intensities between $I$ and $I+dI$. We assume that the laser beam has Lorentzian spatial distribution of the intensity along the propagation direction and Gaussian intensity profile in the transverse direction (see, e.g., Refs.~\cite{SiegmanBook,Augst1991,Lin2018}). 
The focal volume element for such a beam is given by \cite{Morishita2007}:
\begin{equation}
\label{focvol}
\left(-\frac{\partial V}{\partial I}\right)dI\sim\frac{I_0}{I}\left(\frac{I_0}{I}+2\right)\sqrt{\frac{I_0}{I}-1}dI.
\end{equation}
Obviously, the calculation of the focal-volume averaged distribution requires a number of TDSE solutions for different intensities $I<I_{0}$, and therefore is computationally demanding. For this reason, we calculate only $N_a=100$ focal volume averaged PMD's for random internuclear distances $R_k\in\left[1.0, 8.0\right]$~a.u. and peak intensities $I_{k}\in\left[1.0, 4.0\right]\times 10^{14}$ W/cm$^2$ $\left(k=1,...,N_a\right)$.

Figure \ref{fig4}~(b) illustrates the performance of the neural network on this test set. It is seen that the CNN does not perform well for the averaged PMD's. The MAE on this test set reaches the value of 0.83~a.u. However, it is seen that the neural network works relatively well for PMD's that correspond to the internuclear distances less than 5.0~a.u. Indeed, the MAE calculated for the focal-volume averaged PMD's with $R<5.0$~a.u. is equal to 0.24~a.u. which is almost $3.5$ times less than the MAE for the whole averaged set. The unsatisfactory performance of the CNN for focal averaged PMD's with $R>5.0$~a.u. can be understood from a close inspection of Figs.~\ref{fig2} (a)-(e). It is seen that the averaged distributions for $R=5.0$~a.u. and $R=6.0$~a.u. shown in Figs.~\ref{fig2}~(d) and \ref{fig2}~(f), respectively, are similar to each other, especially in their low-energy parts, i.e., for $0<k_{x}<\sqrt{2U_{p}}$ and $\left|k_y\right|<\sqrt{U_p}$. Simultaneously, these distributions are not too similar to their counterparts calculated for fixed laser intensity [cf. Figs.~\ref{fig2}~(c) and \ref{fig2}~(d), as well as Figs.~\ref{fig2}~(e) and \ref{fig2}~(f)]. In contrast to this, the averaged distributions corresponding to smaller values of $R$ resemble the PMD's for same internuclear distances and fixed intensities [cf. Figs.~\ref{fig2}~(a) and \ref{fig2}~(b)]. All this explains why the CNN trained with the distributions for fixed intensities underestimates large internuclear distances by treating them as $R\lesssim5.0$~a.u. 

In order to understand whether the internuclear distance can be reliably retrieved from the focal averaged momentum distributions using deep learning, we train another CNN. This second neural network has the same architecture as the first one (see Fig.~\ref{fig1}), but it is trained on a set of averaged PMD's. A set of $N_a=100$ distributions is too small to train a neural network. Therefore, the data set should be augmented. To this end, we apply 2D interpolation on an irregular grid (see, e.g., Ref.~\cite{Press2007}) in the $\left(R, I_0\right)$ plane formed by the $N_a$ points. As a result, we produce a set of $6000$ focal averaged electron momentum distributions and use them to train our new CNN. In order to have a test set independent on the initial $N_a$ focal averaged momentum distributions, we produce another $N_a=100$ of averaged PMD's for random internuclear distances and peak laser intensities using direct numerical solution of the TDSE and Eq.~(\ref{aver}). We find that the CNN trained on the set of intensity averaged PMD's shows a rather good performance, see Fig.~\ref{fig4}~(c). Despite the presence of a few outliers, the MAE on the independent test set is about 0.14~a.u. This result clearly shows that neural networks can be used to retrieve the unknown internuclear distance from a given electron momentum distribution even if the latter is affected by focal averaging. 

\section{Conclusions and Outlook}

In conclusion, we have investigated the capabilities of deep learning for retrieval of the internuclear distance in the H$_2^{+}$ molecule from a given 2D electron momentum distribution generated by a strong laser pulse. We have shown that the neural network trained on only a few thousand images is able to predict the internuclear distance with a MAE less than $0.1$~a.u. Furthermore, we have studied the effect of focal averaging on the retrieval of the internuclear distance with a neural network. It is shown that the CNN trained on a set of focal averaged distributions also performs well.

The electron momentum distributions are sensitive not only to intensity fluctuations, but also to the changes of other laser parameters. For short laser pulses the variations of the CEP can change the resulting PMD's significantly. Therefore, the effect of the CEP on the retrieval of the internuclear distance needs to be studied. Moreover, it is of interest to look ``inside" the CNN and analyze what features of the holographic structures allows the network to classify the images according to the corresponding values of $R$. This can be achieved by inspection of the transformations of the input PMD by different layers of the neural network. These questions will be the subject of further studies. Progress in these directions is important for the development of the SFPH and for the whole field of time-resolved molecular imaging.

\section{Acknowledgements}

We are grateful to Simon Brennecke, Florian Oppermann, and Shengjun Yue for continued interest to this work and stimulating discussions. This work was supported by the Deutsche Forschungsgemeinschaft (Grant No. SH 1145/1-2).

\end{document}